\documentclass{PoS}
\usepackage{cite}
\usepackage{amsmath}
\usepackage{amssymb}
\usepackage{graphicx}
\newcommand{\half}{\frac{1}{2}}

\title{Majorana neutrino masses in gauge-Higgs unification}

\ShortTitle{Majorana neutrino masses in gauge-Higgs unification}

\author{\speaker{K. Hasegawa }\\
        Department of Physics, Kobe University, Kobe 657-8501, Japan.\\
        E-mail: \email{kouhei@phys.sci.kobe-u.ac.jp}}

\abstract{
The theory that the extra-space component of the gauge field is identified
with the standard model Higgs boson, is called the gauge-Higgs unification\,(GHU)
scenario. We examine how the small neutrino masses are
naturally generated in the GHU framework. We find out two model classes
where the following matter multiplets are introduced\,:
1.\,adjoint\,rep. lepton $\Psi_{A}$, 
2.\,fundamental rep.\,lepton $\Psi_{F}$ and scalar $\Sigma_{F}$.
We present a concrete model in each class. At the model in class 1,
the neutrino masses are generated by the admixture of the seesaw mechanism
type-I and -III. At the model in class 2, the masses are generated by the 
inverse seesaw mechanism. 
}

\FullConference{Corfu Summer Institute 2018 "School and Workshops on Elementary Particle Physics and Gravity"\\
		(CORFU2018)\\
		31 August - 28 September, 2018\\
		Corfu, Greece}

\begin{document}
\section{Introduction}
The Higgs boson was discovered at the Large Hadron Collider\,(LHC) in the year 2012.
The present best-fit values of the mass is 
$M_{H}^{ATLAS}=124.97\pm 0.24 \ \mbox{GeV}$
from the ATLAS Collaboration \cite{Aaboud:2018wps}, 
and  
$M_{H}^{CMS}=125.26  \pm 0.21 \ \mbox{GeV}$
from the CMS Collaboration \cite{Sirunyan:2017exp}.
The gauge coupling constants with the W/Z bosons and the Yukawa couplings
with the heavy quarks and leptons are also confirmed in certain accuracy.
These experimental results about the Higgs boson are well consistent with 
the standard model\,(SM). As far as the Higgs mass, it can not be predicted
in the framework of SM. The mass is a free parameter at the Higgs potential
which leads to the gauge hierarchy problem. 
When we focus on this feature of the SM, we may have the hope that the models 
beyond SM possess the prediction ability for the observed Higgs mass value.
In our knowledge, the beyond standard model\,(BSM) scenarios which can predict 
the Higgs mass is
not so many. One of them is the gauge-Higgs unification\,(GHU) scenario,
where the Higgs boson is identified with the extra-dimensional components
of the higher dimensional gauge fields 
\cite{Manton:1979kb,Hosotani:1983xw,
Hosotani:1983vn,Hosotani:1988bm}.
Since the induced Higgs potential is controlled by the gauge symmetry, 
the Higgs mass is predicted.
At five dimension\,(5D) space-time, the quantum correction to the Higgs mass
is proved to be ultraviolet finite, which gives one solution to
the gauge hierarchy problem \cite{Hatanaka:1998yp}. 
At six dimension\,(6D), the correction diverges,
but after the renormalization, the mass is automatically controlled to the
same order with the W/Z boson masses, which is good agreement with
the experimental value 125\,GeV at the factor two level accuracy
\cite{Manton:1979kb,Antoniadis:2001cv,Csaki:2002ur,Hasegawa:2015vqa}.
In these senses the GHU is a nice candidate for BSM.

At the present stage probably the only one experimental fact for which
the SM must be extended, is the small neutrino masses. The upper bound
for the absolute values is the order of eV, and the bound will become 
0.2\,eV in near future \cite{Drexlin:2013lha}.
The neutrino oscillation experiments revealed the non-zero mass squared
differences at the solar and atmospheric experiments as 
$
(\Delta m_{\odot}^{2}, \Delta m_{atm}^{2}) \simeq
( 10^{-4}\, , 10^{-3}\,)\,\mbox{eV}^{2}
$\,\cite{Tanabashi:2018oca}\,.
These neutrino masses are much smaller than the other SM fermion masses
by the several orders of magnitude.
Then it is requested that the SM is extended to induce such small neutrino
masses in a natural way. Among the SM matter fermions only the neutrinos are
allowed to have the Majorana-type mass term due to the neutral electric charge.
The mechanism to induce the small Majorana neutrino masses in a natural way
is known as the seesaw mechanism
\cite{Yanagida:1979as,GellMann:1980vs,Mohapatra:1979ia}.
For the implementation of the mechanism to the concrete BSM models,
three main types are suggested as 
type-I 
\cite{Yanagida:1979as,
GellMann:1980vs, 
Mohapatra:1979ia},
type-II
\cite{Magg:1980ut,
Cheng:1980qt,
Schechter:1980gr,
Lazarides:1980nt,
Mohapatra:1980yp}, 
and type-III
\cite{Foot:1988aq}
.
When we explore the GHU as candidate for BSM, it should be clarified how
the small Majorana neutrino masses are naturally generated at the GHU framework.
This is the aim of our paper \cite{Hasegawa:2018jze} and this presentation.
\section{Operator Analysis}
Our strategy to find out the way of the neutrino mass extension in GHU is the
following. The specific models in GHU at higher dimensional space-time are
assumed to induce the SM plus the Majorana neutrino masses at four dimension\,(4D)
space-time after Kaluza-Klein expansion.
At 4D there appears a well-known effective operator to induce the Majorana neutrino 
mass. We examine how the effective operator at 4D can be embedded in the GHU 
framework at the higher-dimension. At 4D the effective operator is written as
\begin{equation}
\frac{1}{M}(\overline{L^{c}}\,\epsilon\,\Phi)(\Phi^{t}\,\epsilon\,L)
\simeq \frac{1}{M}(L\Phi)^{2}\,,
\label{eq1}
\end{equation}
where $L$ and $\Phi$ are the lepton and Higgs doublets with 
the anti-symmetric matrix, $\epsilon_{12}=-\epsilon_{12}=1$\,.
The $M$ is the mass scale of a new field for BSM extension, which is assumed to
be much higher than the electroweak scale.
This operator is the mass dimension five and breaks the lepton number explicitly.
When the Higgs boson gets the vacuum expectation value\,(VEV), 
this operator induces the small Majorana mass term,  
$
\frac{v^{2}}{M^{2}}\,\,\overline{(\nu_{\mbox{{\tiny L}}})^{c}}\,(\nu_{\mbox{{\tiny L}}})
$
, with the VEV $v=246$\,GeV\,. 
At all the seesaw types-I,\,II, and III, the neutrino mass generation is universally
described in term of the dimension five operator.
At the type-I/III, introducing $SU(2)_{L}$-singlet/triplet right-handed neutrino,
the Majorana mass term breaks the lepton number.
At the type-II, introducing $SU(2)_{L}$-triplet complex scalar $\Delta$, the two
interactions,
$\overline{(L)^{c}}\Delta L$ and $\Phi^{T}\Delta \Phi$
, break it.

Then we here search for the possibility to embed the operator in Eq.(\ref{eq1})
into the GHU framework.
We assume that the space-time is the product of the Minkowski four dimension and
any compact $d$-dimension space,
$M^{4}\times C^{\,d}$, 
and the gauge group is the G = SU(n) series.
In the GHU, the lepton doublet $L$ is embedded in a multiplet $\Psi$ of the 
gauge group G, and the Higgs doublet $\phi$ is embedded in the extra dimension
component of the gauge field $A_{y}$, expressed as
\begin{equation}
L \subset \Psi \ \ \ \mbox{and} \ \ \ \Phi \subset A_{y}\,.
\end{equation}
Then the operator $(L\Phi)^{2}$ is embedded as 
$(L\Phi)^{2} \subset (\Psi A_{y})^{2}$\,.
First we examine the case that $\Psi$ is the G-fundamental representation
(G-fund rep.).
In the case, the operator $(\Psi A_{y})^{2}$ can not be formed in any gauge-invariant
combination, because the gauge filed $A_{y}$ belongs to adjoint rep. Taking this lesson,
we can find out the following two possible classes,
\begin{align}
\mbox{Class 1}&:\ \ 
(L\Phi)^{2} \subset  (\Psi_{A} A_{y})^{2} \subset
Tr[
(D_{N}\Psi_{A})(D^{N}\Psi_{A})
]
\nonumber\\
&\hspace{20mm}\mbox{with} \ \Psi_{A}: \mbox{G--adjoint rep.}\,,
\label{cl1}
\\
\mbox{Class 2}&:\ \  
(L\Phi)^{2} \subset   (\langle s^{\dagger} \rangle
 L\Phi )^{2} \subset  
(
\Sigma_{F}^{\dagger}
\Psi_{F} A_{y})^{2} \subset  (\Sigma_{F}^{\dagger}D_{N} \Psi_{F})^{2}
\nonumber\\
&\hspace{20mm}\mbox{with} \ \Psi_{F},\Sigma_{F} : \mbox{G--fund rep.}\,,
\label{cl2}
\end{align}
where the index $N$ runs over 5D as $N=(\mu,y)$, and the $D_{N}$
is covariant derivative.
In Eq.(\ref{cl2}) 
the G-fund rep. scalar field, $\Sigma_{F}$, contains the SM-singlet $s$
as $s \subset \Sigma_{F}$.
We present the concrete models in the classes 1 and 2 at the following sections
3 and 4, respectively.
\section{Model 1}
We present a concrete model in the class 1 in Eq.(\ref{cl1}).
The space-time is set to 
$M^{4} \times S^{1}/Z_{2}$, 
and the gauge group is SU(3).
Then we introduce one SU(3)-adjoint(octet) matter fermion as
$\Psi=\Psi^{a}t^{a}$,
where $\Psi^{a}$ is complex field and $t^{a}$ is the Gell-Mann matrices.
The adjoint fermion part of the action is given as
\begin{equation}
S=\int d^{4}x \int_{-\pi R}^{\pi R} dy \
 Tr\,\left[\,
 2\bar{\Psi} i \gamma^{N}D_{N}\Psi
 - M(\bar{\Psi}\gamma^{5}\Psi^{c} 
-\overline{\Psi^{c}}\,\gamma^{5}\Psi 
) 
 \right]\,,
\label{action}
\end{equation}
where the $\gamma^{N}$ is the gamma matrices in 5D as 
$
\gamma^{N}=(\gamma^{\mu},\gamma^{y})=(\gamma^{\mu},i\gamma^{5})
$,
with 4D gamma matrices $\gamma^{\mu}$ and the chiral matrix $\gamma^{5}$.
The covariant derivative is written as 
$
D_{N}\Psi=\partial_{N}\Psi - ig[A_{N},\Psi]
$
with the gauge coupling $g$.
The charge conjugation $\Psi^{c}$ is defined in the same manner as the 4D case,
$
\Psi^{c}
=(\Psi^{a})^{c}\,t^{a}
=\gamma^{2}(\Psi^{a})^{\ast}\,t^{a}
$.
In this model the bulk mass term, 
$
Tr\left[
M\,\bar{\Psi}\gamma^{5}\Psi^{c} 
\right]\,+\,h.c.
$
, breaks the lepton number explicitly.
This mass term is easily shown to be gauge invariant of the group SU(3),
and also Lorentz
invariant in 5D space-time.
It is noticed that at 5D space-time there is no Majorana spinor nor Majorana
mass term. But the Majorana-like mass term in Eq.(\ref{action}) is sufficient to violate
the lepton number and to induce the Majorana mass terms at 4D.
One better explanation of the mass term is to start from 6D,
where the symplectic Majorana spinor exists
\cite{Kugo:1982bn,Mirabelli:1997aj}.
Through the naive dimensional reduction from 6D to 5D, the symplectic
Majorana mass term induces the mass term in Eq.(\ref{action}).

Next we see the zero mode sector after the Kaluza-Klein expansion.
The pattern of the gauge symmetry breaking is same as the original version of
this model \cite{Kubo:2001zc}, which is summarized as
\begin{equation}
\mbox{SU}(3) \   \xrightarrow[S^{1}/Z_{2}]{P=diag(1,1,-1)}
  \  \mbox{SU}(2)_{L}  \times  \mbox{U}(1)_{Y} 
 \ \xrightarrow[\langle A_{y} \rangle =v \cdot t^{6}]{} 
\mbox{U}(1)_{em}\,. 
\end{equation}
The non-trivial $Z_{2}$-parity assignment, $P=(+,+,-)$, breaks the SU(3) to
$SU(2)_{L} \times U(1)_{Y}$. Then the VEV of $A_{y}$ is assumed to be
$
\langle A_{y} \rangle =v\,t^{6}
$, which breaks $SU(2)_{L} \times U(1)_{Y}$ to $U(1)_{em}$.
The orbifolding condition and the VEV $\langle A_{y} \rangle$ determine 
the remaining 
zero mode components in the gauge multiplets and the quantum number of the
SM gauge group. The zero modes of the gauge fields $(A_{\mu},A_{y})$
are determined as
\renewcommand{\arraystretch}{1.2}
{\small
$$
 A_{\mu}^{(0)}=
 \frac{1}{\sqrt{2}}
\left(
 \begin{array}{c|c|c}
\gamma \sqrt{2/3} &  W^{+} & 0   \\ \hline
 W^{-} &-Z/\sqrt{2}-\gamma/\sqrt{6}  &  0  \\ \hline
0  & 0   &Z/\sqrt{2}-\gamma/\sqrt{6}\\
 \end{array}
 \right)\,, \ \ \    
 A_{y}^{(0)}
 =
 \frac{1}{\sqrt{2}}
 \left(
 \begin{array}{c|c|c}
 0  & 0  & \phi^{+}   \\ \hline
 0  & 0  & \phi^{0}   \\ \hline
 \phi^{-} & \phi^{0\ast} &  0  \\
 \end{array}
 \right)\,.
$$
\renewcommand{\arraystretch}{1}
}
The chiral zero modes of the adjoint fermion $\Psi$ appears as
{\footnotesize
\renewcommand{\arraystretch}{1.3}
$$
 \Psi_{L}^{(0)}=
 \frac{1}{\sqrt{2}}
 \left(
 \begin{array}{c|c|c}
 0  & 0   & \tilde{e}^{+}   \\ \hline
 0  & 0   & \tilde{\nu}   \\ \hline
 e^{-}  & \nu  & 0 \\
 \end{array}
 \right)_{L}\,, 
\ \  \  
\Psi_{R}^{(0)}
 =
\frac{1}{\sqrt{2}}
 \left(
 \begin{array}{c|c|c}
 N^{\gamma}\sqrt{2/3}& \tilde{E}^{+}  & 0   \\ \hline
 E^{-}  & -N^{z}/\sqrt{2}-N^{\gamma}/\sqrt{6} & 0   \\ \hline
 0  & 0   &  N^{z}/\sqrt{2}-N^{\gamma}/\sqrt{6}  \\
 \end{array}
 \right)_{R}\,,
$$
\renewcommand{\arraystretch}{1}
}
with $\gamma^{5}\Psi_{R/L}^{(0)}=\mp \Psi_{R/L}^{(0)}
$.
The neutral charge components $(N^{z},N^{\gamma})$, are defined as
\begin{equation}
 \left(
    \begin{array}{c}
       N^{Z} \\
       N^{\gamma}
    \end{array}
  \right)_{R}=
  \left(
    \begin{array}{cc}
      \cos \theta_{W} &  -\sin \theta_{W} \\
     \sin \theta_{W} &  \cos \theta_{W} 
    \end{array}
  \right)\left(
    \begin{array}{c}
      \Psi^{a=3} \\
      \Psi^{a=8}
\end{array}
  \right)_{R}.
\end{equation}
with the angle $\theta_{W}=60^{\circ}$.
Here the component fields,
$\Psi^{a=3}$ and $\Psi^{a=8}$ in $\Psi=\Psi^{a}t^{a}$,
belong to the $SU(2)_{L}$-triplet and -singlet, respectively.

\begin{figure}[t]
\begin{center}
\includegraphics[width=7.0cm]{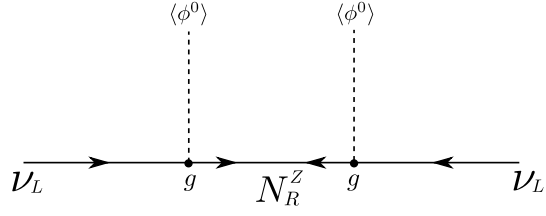}
\end{center} 
\caption{ 
The Feynman diagram which represents the neutrino mass generation
in the model 1 is shown. The right-handed neutrino $N_{R}^{Z}$ is 
the mixing state of the $SU(2)_{L}$-1 and -3 representations. 
\label{fig1}}
\end{figure}

Finally we see how the neutrino masses are generated in this model.
We extract the Dirac neutrino mass terms induced by the Higgs
VEV, $\langle \phi^{0} \rangle=v/\sqrt{2}$\,. The Yukawa interaction with
$A_{y}^{(0)}$ is contained in the action in Eq.(\ref{action}) as
\begin{align}
{\cal L}_{yukawa} 
&= 2g\,Tr[\,\bar{\Psi} \gamma^{y}\,[A_{y},\Psi]\,]\nonumber\\
&= 2ig\,  Tr[\bar{\Psi}_{L} \,[A_{y},\Psi_{R}]\,] + h.c. 
\end{align}
In the Yukawa interaction, the contribution including both of 
the Higgs VEV $\langle \phi^{0} \rangle$ 
and the neutrinos, is the only following one,
\begin{equation}
{\cal L}_{yukawa} 
\supset
2ig\,Tr \left[
\overline{\Psi_{L}^{7}}t^{7}
\,[
\sqrt{2}\langle \phi^{0}\rangle t^{6},
N_{R}^{Z}t_{Z}
]\,
\right]
+ h.c.,
\end{equation}
with 
$
t_{Z}
=(t^{3}-\sqrt{3}t^{8})/2
$\,.
This contribution is calculated as
\begin{align}
{\cal L}^{Dirac}_{neutrino} 
&=ig \langle \phi^{0} \rangle  \,  
\bigl(\,\overline{\nu_{L}} - \overline{\tilde{\nu}_{L}} 
\,\bigr)N_{R}^{Z}
 +h.c.,
\label{dirac}
\end{align}
where we use the definition,  
$
\Psi_{L}^{7}=i(\,\nu_{L} - \tilde{\nu}_{L} 
\,)/\sqrt{2}
$.
Next we extract the neutrino mass terms from 
the Majorana-like bulk mass term in Eq.(\ref{action}) as
\begin{equation}
{\cal L}^{Majorana}_{neutrino} 
= -M \,   \left(\overline{\nu_{L}}  (\tilde{\nu}_{L})^{c}
- \half\overline{N_{R}^{Z}} (N_{R}^{Z})^{c}
 - \half\overline{N_{R}^{\,\gamma}} (N_{R}^{\,\gamma})^{c}
\right)
+h.c. 
\label{majorana}
\end{equation}
In Eq.(\ref{dirac}), 
there exists the Dirac mass term formed from the left-handed spinor,
$(\nu_{L} - \tilde{\nu}_{L})$,
and the right-handed spinor, 
$N_{R}^{Z}$.
In Eq.(\ref{majorana}), 
there is the Majorana mass term of $N_{R}^{Z}$. 
When the mass $M$ is taken to be 
much higher than the Dirac mass $g \langle \phi^{0} \rangle$,  
the right-handed neutrino $N_{R}^{Z}$ is decoupled and then the seesaw 
mechanism seems to work. 
But there is one difference from the usual seesaw setup,
which is the existence of the Dirac mass term in Eq.(\ref{majorana}), 
formed from the left-handed $\nu_{L}$ and the right-handed $(\tilde{\nu}_{L})^{c}$. 
When the mass parameter $M$ is much higher than $g \langle \phi^{0} \rangle$,
both of the Weyl spinors $\nu_{L}$ and $(\tilde{\nu}_{L})^{c}$,
are also decoupled. To decouple the only $\tilde{\nu}_{L}$,
we introduce the brane mass term as
\begin{equation}
{\cal L}_{brane}= 
\delta(y)\,M_{brane}\,\overline{(\,e_{b}^{+},\, \nu_{b})_{R}}
\, \left(\begin{array}{c}
\tilde{e}^{\,+}\\
\tilde{\nu}  
\end{array}
\right)_{L}\,, 
\end{equation}
where the $SU(2)_{L}$-doublet and right-handed lepton 
$(\,e_{b}^{+},\, \nu_{b})_{R}$
is introduced inside the brane at $y=0$.
Then we assume the hierarchy,
$ 
M_{brane} \gg M \gg  g \langle \phi^{0} \rangle
$.
In this case, $\tilde{\nu}_{L}$ is decoupled and the seesaw mechanism works.
The seesaw diagram is shown in Fig.\ref{fig1}.
Since the $N_{R}^{Z}$ is the mixing state of the $SU(2)_{L}$-1 and -3
representations,
this seesaw mechanism is operating as the admixture of the seesaw type-I and -III 
in terminology of the usual 4D case.
The small mass eigenvalue for $\nu_{L}$ is realized as
\begin{equation}
m_{\nu}=\frac{(g \langle \phi^{0} \rangle)^{2}}{M}
\simeq 
\frac{M_{W}^{2}}{M}\,,
\end{equation}
where $M_{W}$ is the W boson mass.  
\section{Model 2}
\begin{figure}[t]
\begin{center}
\includegraphics[width=14cm]{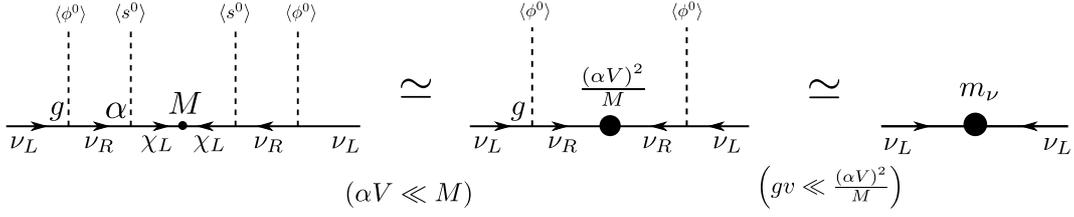}
\end{center} 
\caption{ 
The diagrams which represent the neutrino mass generation
in the model 2 is shown.
\label{fig2}}
\end{figure}

We here briefly present a concrete model in class 2 in Eq.(\ref{cl2}).
To induce the mass dimension seven effective operator, 
$(\Sigma_{F}^{\dagger}D_{M} \Psi_{F})^{2}$,
the fermion $\Psi_{F}$ and the scalar $\Sigma_{F}$ 
of the G=SU(n)-fund rep. must be introduced.
At the G=SU(3) model on the space-time $M^{4} \times S^{1}/Z_{2}$, 
the fermions in fund rep. have the fractional charges as 
$(2/3,-1/3,-1/3)$,
which are suitable for the quarks. For the integer charges of the lepton,
the SU(4) model on $M^{4} \times S^{1}/Z_{2}$ is found to be possible.
Then we introduce the SU(4)-fund rep. fermion $\Psi_{F}$ 
and scalar $\Sigma_{F}$.
In order to violate the lepton number, we introduce the SU(4)-singlet
fermion $\chi$ with the Majorana-like mass term,
$
\frac{M}{2}\bar{ \chi}\gamma^{5}\chi^{c} + h.c.
$
which has the same Lorentz structure as in Eq.(\ref{action}).
The Yukawa interaction
$
\alpha \bar{\Psi}_{F} \Sigma_{F} \chi
$
is also introduced  
with the Yuakwa coupling $\alpha$.
The gauge symmetry breaking in this model is following.
The orbifold condition of the space $S^{1}/Z_{2}$ on two 
fixed points is given as
$
P=(++,++,+-,--)
$.
Both of the orbifold condition and the VEV 
$\langle \Sigma_{F} \rangle$, break as 
$
\mbox{SU}(4) \   
\to  \  \mbox{SU}(2)_{L}  \times  \mbox{U}(1)_{Y} 
$,
which is further broken to  
$
\mbox{U}(1)_{em}
$
by the VEV $\langle A_{y} \rangle$.
The zero mode components of
$\Psi_{F}$, $\Sigma_{F}$, $\chi$, and $A_{y}$,
are shown as
\begin{equation}
\Psi_{F}^{(0)} = 
 \left(
    \begin{array}{c}
      \nu_{L} \\
      e^{-}_{L} \\ 
      0  \\ \hline 
     \nu_{R} \\ 
    \end{array}
  \right)\,, \ \    
\Sigma_{F}^{(0)} = 
 \left(
    \begin{array}{c}
      0 \\
      0 \\ 
      0 \\ \hline 
      s^{0} \\ 
    \end{array}
  \right), \ \
\chi^{(0)}=\chi_{L}^{0}\,, \ \ 
A_{y}^{(0)} = 
\left(
    \begin{array}{ccc|c}
      0 & 0 & 0 &\phi^{0} \\
      0 & 0 & 0  & \phi^{-} \\ 
      0 & 0 & 0  & 0 \\ \hline 
      \phi^{0 \ast} &\phi^{+} & 0 & 0 \\ 
    \end{array}
  \right)\,. 
\end{equation}
Then we see the neutrino mass generation. The dimension seven operator
$(\Sigma_{F}^{\dagger}D_{M} \Psi_{F})^{2}$, 
contains the operator 
$
(s^{0 \ast}\,\phi^{0}\,\nu_{L})^{2}
$
,
which induces the Majorana mass term for the SU(2)-doublet neutrino
$\nu_{L}$
with the VEVs, $\langle s^{0} \rangle=V/\sqrt{2}$ 
and $\langle \phi^{0} \rangle=v/\sqrt{2}$\,.
Guided by the dimension seven operator,
we can find the corresponding diagram shown at most left in 
Fig.\ref{fig2}.
For the sufficiently large $M$ as $\alpha V \ll M$,
the Majorana mass of the $\mbox{SU}(2)_{L}$-singlet neutrino 
$\nu_{R}$ is induced as 
$
(\alpha V)^{2}/M
$,
shown at the center in Fig.2.
Further, when the hierarchy 
$
gv \ll (\alpha V)^{2}/M
$ 
is assumed, the small Majorana mass term of $\nu_{L}$ is induced as
\begin{equation}
m_{\nu}=\frac
{
(gv)^{2}
}
{
\ \  \frac{(\alpha V)^{2}}{M}
\ \ }\,.
\end{equation}
Such two sequences of the seesaw mechanism may be called 
the inverse seesaw
\cite{Mohapatra:1986aw,Mohapatra:1986bd}, 
and the same mechanism appears also at the context of the grand 
GHU\cite{Hosotani:2017hmu}.
The condition 
$
gv \ll (\alpha V)^{2}/M
$, 
is naturally satisfied when the Yukawa coupling $g$, which is nothing but 
the gauge coupling in GHU, is small. If the bulk mass term for
$\Psi_{F}$
with the sign function,
$
\epsilon(y)M_{F}\bar{\Psi}_{F}\Psi_{F}
$,
is added, the Yukawa coupling is suppressed as 
$
f=g(\pi R M_{F}) e^{-\pi R M_{F}}
$ 
with the radius $R$ of $S^{1}$ \cite{Hasegawa:2012sy}. 
This implementation is adapted in our paper
\cite{Hasegawa:2018jze}.

\section{Summary}
The GHU scenario is a nice candidate for BSM.
The possible ways how the small Majorana neutrino masses are 
naturally induced in GHU, are examined. Based on the operator analysis, 
our strategy is to find the effective operator 
in the GHU framework, which induces the well-known 
mass dimension five operator $(L\Phi)^{2}$ at 4D.
In the series of the gauge group G=SU(n), we find out two model classes
where the following matter multiplets are introduced\,:
1.\,G-adjoint rep. fermion $\Psi_{A}$, 2.\,G-fund rep. fermion $\Psi_{F}$ and scalar 
$\Sigma_{F}$.
Then we present a concrete model in each class.
In class 1, the SU(3)-adjoint rep.\,lepton $\Psi_{A}$ is introduced with the Majorana-like
mass term, which violate the lepton number.
The small neutrino masses for the active left-handed
neutrino are naturally induced by the mixture of the seesaw 
mechanism type-I and -III.
In class 2, the SU(4)-fund rep.\,lepton $\Psi_{F}$ and scalar $\Sigma_{F}$, are 
introduced. Further the
SU(4)-singlet lepton $\chi$ is introduced with the Majorana mass term which 
breaks the lepton number. The small neutrino masses are realized by the
inverse seesaw mechanism. One of the future plan is the construction
of the models which can predict the mass eigenvalues of three generation neutrinos
and the mixing angles in a natural way. 


\providecommand{\href}[2]{#2}\begingroup\raggedright\endgroup

%

\end{document}